\documentclass[sigconf]{acmart}

\usepackage{amsmath, amsfonts}
\usepackage{algorithmic}
\usepackage{graphicx}
\usepackage[ruled, vlined]{algorithm2e}
\usepackage{textcomp}
\usepackage{multirow}
\usepackage{hyperref}
\usepackage{outlines}
\usepackage{soul}
\usepackage{booktabs}  
\usepackage{subcaption}
\usepackage{microtype}

\SetKwBlock{Repeat}{Repeat until convergence:}{}
\usepackage{newfloat}
\usepackage{listings}
\usepackage{caption}
\usepackage{enumitem}

\newcommand{\revised}[1]{{\textcolor{black}{#1}}}

\copyrightyear{2022} 
\acmYear{2022} 
\setcopyright{rightsretained} 
\acmConference[WPES '22]{Proceedings of the 21st Workshop on Privacy in the Electronic Society}{November 7, 2022}{Los Angeles, CA, USA}
\acmBooktitle{Proceedings of the 21st Workshop on Privacy in the Electronic Society (WPES '22), November 7, 2022, Los Angeles, CA, USA}
\acmDOI{10.1145/3559613.3563201}
\acmISBN{978-1-4503-9873-2/22/11}


\begin{document}

\title{UnSplit: Data-Oblivious Model Inversion, Model Stealing, and Label Inference Attacks Against Split Learning}
    
\author{Ege Erdoğan}
\affiliation{%
  \institution{Koç University}
  \city{\.Istanbul}
  \country{Turkey}
}
\email{eerdogan17@ku.edu.tr}

\author{Alptekin Küpçü}
\affiliation{%
  \institution{Koç University}
  \city{\.Istanbul}
  \country{Turkey}
}
\email{akupcu@ku.edu.tr}

\author{A. Ercüment Çiçek}
\affiliation{%
  \institution{Bilkent University}
  \city{Ankara}
  \country{Turkey}
}
\email{cicek@cs.bilkent.edu.tr}

\begin{abstract}
Training deep neural networks often forces users to work in a distributed or outsourced setting, accompanied with privacy concerns. \textit{Split learning} aims to address this concern by distributing the model among a client and a server. The scheme supposedly provides privacy, since the server cannot see the clients' models and inputs. We show that this is not true via two novel attacks. (1) We show that an honest-but-curious split learning server, equipped only with the knowledge of the client neural network architecture, can recover the input samples and obtain a functionally similar model to the client model, without being detected. (2) We show that if the client keeps hidden only the output layer of the model to "protect" the private labels, the honest-but-curious server can infer the labels with perfect accuracy. We test our attacks using various benchmark datasets and against proposed privacy-enhancing extensions to split learning. Our results show that plaintext split learning can pose serious risks, ranging from data (input) privacy to intellectual property (model parameters), and provide no more than a false sense of security.
\end{abstract}

\maketitle 


\section{Introduction}

\begin{figure*}[t!]
    \centering
    \begin{subfigure}{0.48\linewidth}
        \includegraphics[width=\textwidth]{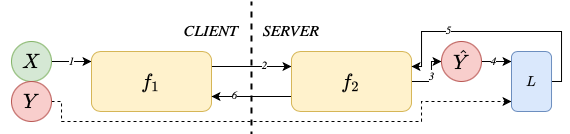}
        \caption{Training examples and labels at the client.}
        \label{fig:splitnn_client_data}
    \end{subfigure}
    
    \begin{subfigure}{0.48\linewidth}
        \includegraphics[width=\textwidth]{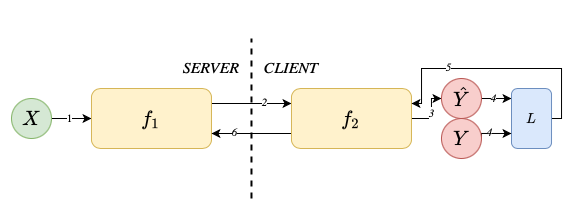}
        \caption{Training examples and labels are split between the client and the server.}
        \label{fig:splitnn_server_data}
    \end{subfigure}
    \hspace{0.2cm}
    \begin{subfigure}{0.48\linewidth}
        \centering
        \includegraphics[width=\textwidth]{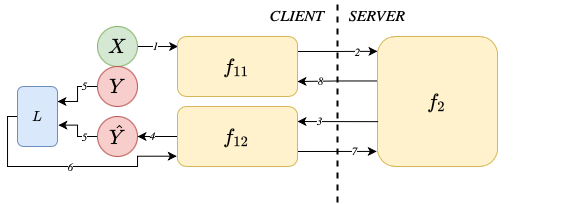}
        \caption{Training examples and labels stored only at the client.}
        \label{fig:splitnn_client_label}
    \end{subfigure}
    
    \caption{Three possible SplitNN setups. A client and a server train a model with the dataset containing examples $X$ and labels $Y$, where $\hat{Y}$ stands for the model's predictions, and $L$ for the loss function. The numbers on the edges denote the steps of computation in order.}
    \label{fig:splitnn_setups}
\end{figure*}

There has been two trends that has fueled the recent growth in the interest towards deep neural networks (DNNs): increasing computing power, and availability of large datasets. Training a DNN with millions or billions of parameters is an expensive task that requires significant computing power. It is also known that having access to large high-quality training data alone is generally enough to increase a model's performance \citep{halevy_unreasonable_2009}. Due to these reasons, distributed and outsourced approaches to model training that split the data storage and computation loads among multiple nodes have attracted attention.

\textit{Federated learning} \citep{bonawitz_towards_2019, konecny_federated_2016} and \textit{split learning} (SplitNN) \citep{gupta_distributed_2018, vepakomma_split_2018, vepakomma_no_2018} are two distributed deep learning frameworks proposed to further the two trends described above, by (i) enabling more efficient training of DNNs on devices with limited capabilities (e.g. smartphones), and 
(ii) allowing multiple data holders to train a DNN without sharing, but utilizing, their aggregate data. However, various studies have shown that these techniques leak information \citep{calandrino_you_2011, he_model_2019, li_label_2021, zhao_idlg_2020, zhu_deep_2019, fredrikson_model_2015}. Especially for strictly-regulated fields such as healthcare, it is of critical importance to ensure that distributed neural network training is also privacy-preserving.

The framework we focus in this paper, \textbf{split learning} or \textbf{SplitNN} \citep{gupta_distributed_2018, vepakomma_split_2018}, allows one or more clients to train a DNN by splitting the DNN so that the first few layers are computed at the client(s), and the rest at a central server. A client shares its final layer's output, called \textit{smashed data}, rather than its private data. We detail this in Section \ref{background}. Compared to other similar frameworks, SplitNN stands out as being more efficient \citep{vepakomma_no_2018}. 

\textbf{Contributions.} In this paper, we present UnSplit: a suite of two novel attacks against SplitNN that effectively "unsplits" the split. Compared to previous similar attacks, our attacks work equally-well with the least amount of client-side knowledge needed by the attacker server. The two attacks can be summarized as follows:
\begin{itemize}[nosep]
    \item The first attack allows a SplitNN server to recover the client's inputs given to the model, while also obtaining a functionally similar model to the client model. We assume that the attacker knows only the architecture of the client model. With this threat model, the attack surface consists only of the clients' smashed data. 
    \item The second attack is a label inference attack that allows an honest-but-curious SplitNN server to infer the supposedly protected labels with perfect accuracy, assuming that the client only computes the output layer locally. While this is a simplistic assumption, the effectiveness and potential consequences of the attack deems it worthy of discussion.
\end{itemize}
Although we focus on the single-client setting for clarity, our attacks generalize to multi-client settings without any modification. We further detail this after presenting our attacks. 

In our attacks, the server is an \textbf{honest-but-curious} attacker: it acts according to the SplitNN protocol, but performs the attack in the background. Such attacks cannot be detected by a client, since the protocol is followed as expected. Thus, our adversary is very weak, requiring minimal assumptions, but the results of our attack are potentially devastating regarding privacy.

\revised{There has been past work demonstrating ways data privacy and model confidentiality can be violated in collaborative learning setups \cite{he_model_2019, pasquini_unleashing_2021, zhu_deep_2019, zhao_idlg_2020, zhang_secret_2020, salem_updates-leak_2019, fredrikson_model_2015}. UnSplit's novelty stems from its symbiotic combination of model stealing and model inversion in a limited threat model within the context of split learning. By combining these two attacks, the adversary obtains more effective results compared to earlier attacks of the same nature as we will present in the following sections.}

Our attacks introduce the adversarial goal of model stealing into the SplitNN setting, and demonstrate that attempting model stealing along model inversion improves the quality of the recovered inputs, thus posing a significant risk to both input privacy and model confidentiality.

The code for our approach is available at \url{https://github.com/ege-erdogan/unsplit}.

\section{Background and Related Work}\label{background}

SplitNN \citep{gupta_distributed_2018, vepakomma_split_2018} is a distributed deep learning framework that enables multiple data holders and a central server to collaboratively train a DNN, without any data sharing between the participants. Such distributed methods can provide substantial benefits in areas such as healthcare, where data holders (e.g. hospitals, clinics) are prohibited from sharing their data due to regulations such as HIPAA \citep{annas_hipaa_2003, mercuri_hipaa-potamus_2004}.

SplitNN's main idea is to split and allot a DNN among multiple parties. Figure \ref{fig:splitnn_setups} displays potential setups of the SplitNN protocol. In its simplest setting (Figure \ref{fig:splitnn_client_data}), SplitNN involves a single data holder (client) and a server. The client computes the first few layers of the DNN, and forwards the output, along with the target label, to the server. The server then resumes the computation with the remaining layers. The server then initiates the corresponding backward pass by computing the loss value, and sends the gradient values of its first layer to the client. The client completes the backward pass, following the backpropagation from the received gradients.

The above setting requires the client to share the training labels with the server. We can omit that requirement by further dividing the network into three parts, with the final part being computed at the client (Figure \ref{fig:splitnn_client_label}). The only difference is the addition of one more communication step. Since the loss value is computed at the client, the client does not need to share labels with the server. Alternatively, in another setting that does not require any data sharing, the server can store the training examples and the client can store the labels (Figure \ref{fig:splitnn_server_data}).

Multiple clients can participate in SplitNN by following a turn-based training procedure. Before a client starts its turn, it updates its weights with that of the most recently trained client. This can be achieved either through a central server, or in a peer-to-peer basis between the clients.

\begin{table}[ht]
    \caption{Adversary capabilities for various model inversion attacks. All attacks aim to reconstruct inputs given to the target model. UnSplit aims to steal the model as well. Attacks that directly target SplitNN are marked with an asterisk ($^*$) and are at the bottom half of the table.} 
    
    \tabcolsep=0.05cm
    \centering
    \begin{tabular}{cccccc}
    \toprule
    \multirow{3}{*}{Attack} & \multicolumn{5}{c}{Adversary Knowledge / Capabilities} \\ \cmidrule(r){2-6} 
    & \multicolumn{2}{c}{Target Model} & \multicolumn{2}{c}{Training Data} & Send \\
    & Structure & Params           & Values & Distribution             & Query \\
    \midrule
    \citep{zhang_secret_2020}                       & $\times$ & $\times$ & -  &  - & $\times$ \\ 
    \citep{salem_updates-leak_2019}                 &  - &  - & -  &  - & $\times$ \\ 
    \citep{zhu_deep_2019}                           & $\times$ & $\times$ & -  &  - & $\times$ \\
    \citep{fredrikson_model_2015} White-box         & $\times$ & $\times$ & -  &  - & $\times$ \\
    \citep{fredrikson_model_2015} Black-box         &  - &  - & -  &  - & $\times$ \\ 
    \midrule
    \citep{he_model_2019} White-box*                 & $\times$ & $\times$ & -  &  - & -  \\ 
    \citep{he_model_2019} Black-box*                 &  - &  - & $\times$ & $\times$ & $\times$ \\ 
    \citep{he_model_2019} Query-free*                & $\times$ &  - & $\times$ & $\times$ & -  \\
    \citep{pasquini_unleashing_2021}*            &  - &  - & -  & $\times$ & -  \\ 
    \textbf{UnSplit}*                                            & $\times$ &  - & -  &  - & -  \\
    \bottomrule
    \end{tabular}
    
    \label{tab:attacks}
\end{table}

\subsection{Reducing Information Leakage in SplitNN} 

To minimize the information smashed data leaks, \citep{vepakomma2019reducing} introduce an additional loss function to the SplitNN training procedure: the logarithm of distance correlation (DCOR) \citep{szekely2007measuring} between the inputs and the smashed data. Following the notation in Figure \ref{fig:splitnn_setups}, the overall loss function becomes
\begin{equation}
    \alpha_1 DCOR(X, f_1(X)) + \alpha_2 L(Y, \hat{Y})
\end{equation}
where $\alpha_1$ and $\alpha_2$ are used to control the impact of the distance correlation term. For brevity, we refer the reader to \citep{vepakomma2019reducing} for a detailed theoretical analysis linking invertibility with distance correlation.

\subsection{Model Inversion Attacks}

In a model inversion (MI) attack \citep{wu_methodology_2016, calandrino_you_2011, he_model_2019, li_label_2021, zhao_idlg_2020, zhu_deep_2019, fredrikson_model_2015}, an adversary tries to obtain the inputs of a machine learning model, given access to its output. MI attacks are not unique to federated/split learning setups, but are especially important since data privacy is a primary concern of such setups. 

Table \ref{tab:attacks} summarizes the threat models of various MI attacks. Notice that for most attacks, the attacker can send queries to the target model. However, that is not possible for a SplitNN server since the clients control the inputs given to the model.

An early example of a model inversion attack \citep{fredrikson_model_2015} targets a linear regression model used to adjust medicine doses for patients. Given the machine learning model and some demographic information about a patient, the attack was able to predict the patient's genetic markers used as inputs to the model. 

\subsubsection{Attacks Against Split Learning}

Pasquini et al. \citep{pasquini_unleashing_2021} demonstrated that an honest-but-curious server could obtain the clients' data during training. The attack relies on the server's ability to direct the client towards arbitrary goals, independent of the actual task (e.g., classification). The results of the attack demonstrate that the SplitNN protocol is inherently insecure. However, this attack cannot steal the client model.

The attack, named the \textit{Feature-Space Hijacking Attack} (FSHA), assumes an attacker that has access to a data set $X_{pub}$ which follows a similar distribution with that of the client's training data $X_{priv}$. Briefly, the attacker trains an autoencoder on $X_{pub}$ and directs the client towards outputting values belonging to the same latent space as the encoder part of the autoencoder. Since the decoder essentially knows how to invert the values belonging to that latent space, it is able to invert values received from the client, and obtain the original inputs. The main difference between FSHA and UnSplit is that we do not have any assumptions about the attacker's knowledge of a public data set related to the original task. While such a dataset might be available in certain scenarios, its non-existence makes FSHA infeasible since the attacker cannot train the autoencoder. As another point of difference, FSHA does not attempt to steal the client model.

In a different set of MI attacks targeting collaborative inference systems similar to SplitNN, \citep{he_model_2019} showed under various threat models that it is possible to recover the input fed to a DNN with varying degrees of accuracy. They considered \textit{white-box} scenarios, where the adversary knows the parameters of the DNN, \textit{black-box} scenarios, where the adversary does not know the parameters but can query the DNN, and \textit{query-free} scenarios, where the adversary neither knows the weights of nor can send queries to the DNN, has knowledge about the underlying dataset. The most applicable of these threat models to SplitNN is the query-free setting, since a SplitNN server neither knows the parameters of, nor can send queries to the client model. Moreover, the effectiveness of the black-box and query-free attacks heavily depends on the server's knowledge of the original learning task.

\subsubsection{Attacks in Different Settings}

As shown By Zhu et al. \cite{zhu_deep_2019} and further improved by Zhao et al. \cite{zhao_idlg_2020}, an honest-but-curious federated learning server can recover a training input by optimizing for an input resulting in the same gradient values as the original one. This implies that even if the forward and backward passes are performed on the client side and the server is not given capabilities beyond those required by the protocol, sharing the gradient values can leak information.

\section{Method}\label{method}

\subsection{Threat Models}

For our \textbf{model inversion and stealing attack}, we consider a client and a server running the SplitNN protocol, where for simplicity a DNN $F$ is partitioned into two parameterized functions $f_1$ and $f_2$ such that $F(\theta, x)=f_2(\theta_2, f_1(\theta_1, x))$.

We assume an attacker that knows the model architecture, but not the parameters, of $f_1$. The attacker does not have access to any specific data, and strictly follows the SplitNN protocol. This means that the attacker cannot query the client network, and does not send training updates other than the one required for the original learning task. Whether the model terminates at the client (Figure \ref{fig:splitnn_client_label}) or the server (Figure \ref{fig:splitnn_client_data}) is of no importance to the attacker, since the attacker only needs the smashed data she receives from the client. Thus, we model an \textit{\textbf{honest-but-curious}} attacker, which is a much weaker form compared to a powerful malicious attacker.

The attacker's goals are to recover any input given to the network $F$, and obtain a functionally similar (i.e., similar performance on unseen data) clone $\tilde{f_1}$ of the client network $f_1$. Within this threat model, it is impossible for the clients to distinguish a server launching the attack from one following the protocol.

It is important to note that this is a realistic scenario for SplitNN: a researcher (SplitNN server, controlling model design) and healthcare providers (SplitNN clients) can use SplitNN to train a DNN. 

It might be argued that for model stealing, our adversary receiving smashed data resulting from actual training data (although it does not have access to any specific data) yields it stronger than the traditional black-box adversary (only able to query the model) studied within model stealing attacks. While it is difficult to strictly classify one adversary as stronger,\footnote{As one counterpoint, a black-box adversary might have a practically unlimited query budget, while our adversary works over single examples.} we should stress that such a black-box model is not possible within SplitNN (the attacker cannot query the client model), and that our adversary does not go beyond the capabilities SplitNN provides. Since we are mainly concerned with SplitNN and not a universal model stealing attack, these are reasonable assumptions.

For the \textbf{label inference attack} (Figure \ref{fig:splitnn_server_data}), the same assumptions are valid, implying that the attacker knows how many discrete labels there are. We further assume that training updates are calculated with \textit{stochastic} gradient descent, and that the client model has a depth of one. It is reasonable to expect SplitNN to be used with minimal cost for clients, while "protecting" their data. The severity of the attack's consequences deems it worthy of discussion, and highlights the importance of preventing such use.

\subsection{Model Inversion \& Stealing}

Without any data similar to the training data or the ability to query the client network, the attacker's task is a search over the input- and parameter-spaces. We model the problem as an optimization problem: the attacker tries to find parameters $\tilde{\theta_1}$ and input $\tilde{x}$ to minimize the difference between $\tilde{f_1}(\tilde{\theta_1}, \tilde{x})$ and $f_1(\theta_1, x)$ (Equations \ref{eq:input_objective} and \ref{eq:model_objective}).

The optimization problem described above can be solved with gradient-based methods. However, we have observed in our experiments that performing the updates on the input and parameters simultaneously, in a single gradient update, often does not yield favorable results. Instead, we adopt a "coordinate gradient descent" \citep{wright_coordinate_2015} approach. A coordinate descent involves keeping a subset of the parameters fixed while updating another subset.

In UnSplit, we partition the target values into two sets, following their logical separation: the input values $\tilde{x}$ and the parameter values $\tilde{\theta_1}$. Given client's output $f_1(x)$, the attacker first performs gradient descent updates on the estimated input values $\tilde{x}$, keeping $\tilde{\theta_1}$ fixed, and then repeats the same process by keeping $\tilde{x}$ constant and updating $\tilde{\theta_1}$. Algorithm \ref{alg:model_attack} summarizes the attack.

The attack can be modified to obtain more accurate results by tuning various parameters on different levels. The attacker can set the number of gradient descent steps separately for both $\tilde{x}$ and $\tilde{\theta_1}$, as well as the total number of rounds. The attacker also has control over the partitioning of the search space; it can either divide it into more sub-spaces (e.g. by layers), or merge into a single space.

To begin the \textit{model inversion and stealing attack}, the server randomly initializes a model that has the same architecture with the client model. Then, the attacker defines two objective functions, for the input and parameter updates. We minimize the mean squared error (MSE) for both updates. Note that this is independent of the loss function used for the actual training task. Furthermore, since we are working in the image domain (see Section \ref{experimental_results}), we also add a Total Variation \citep{rudin_nonlinear_1992} term to be minimized, following from the work in \citep{he_model_2019}. Total Variation is a measure of the noise present in an image, and minimizing it results in smoother images. It is defined for an image $x$ as
$$
    \textrm{TV}(x) = \sum_{i,j} \sqrt{|x_{i+1, j} - x_{i,j}|^2 + |x_{i, j+1} - x_{i,j}|^2},
$$
where $i$ and $j$ denote the pixel indices.

We can summarize the attacker's task with Equations \ref{eq:input_objective} and \ref{eq:model_objective}. The coefficient $\lambda$ can be set to modify how much the Total Variation term affects the loss function. 
\begin{equation} \label{eq:input_objective}
    \tilde{x}^* = argmin_{\tilde{x}} \, \textrm{MSE}(\tilde{f_1}(\tilde{\theta_1}, \tilde{x}) , f_1(\theta_1, x)) + \lambda \textrm{TV}(\tilde{x})
\end{equation}
\begin{equation} \label{eq:model_objective}
    \tilde{\theta_1}^* = argmin_{\tilde{\theta_1}} \, \textrm{MSE}(\tilde{f_1}(\tilde{\theta_1}, \tilde{x}) , f_1(\theta_1, x))
\end{equation}

\begin{algorithm}[t!]
\SetAlgoLined
\KwResult{$\tilde{x}^*$ and $\tilde{\theta_1}^*$}
 L: objective function \\
 $x$: training example \\
 $f_1$: client model \\
 $f_2$: server model \\
 $\tilde{f_1}$: randomly initialized copy of the client model \\
 \Repeat{
    $\tilde{x}^* = argmin_{\tilde{x}} \, \textrm{L}(\tilde{f_1}(\tilde{\theta_1}, \tilde{x}) , f_1(\theta_1, x)) + \lambda \textrm{TV}(\tilde{x})$ \\
    $\tilde{\theta_1}^* = argmin_{\tilde{\theta_1}} \, \textrm{L}(\tilde{f_1}(\tilde{\theta_1}, \tilde{x}) , f_1(\theta_1, x))$
 }
 \caption{UnSplit: Model Inversion \& Stealing}
 \label{alg:model_attack}
\end{algorithm}

\begin{algorithm}[t!]
\SetAlgoLined
\KwResult{$\tilde{y}^*$}
 $L$: objective function \\
 $(x,y)$: training examples and labels \\
 $f_1$: server model \\
 $f_2$: client model \\
 $\tilde{f_2}$: randomly initialized copy of the client model \\
 $h = \frac{\partial L(f_2(f_1(x)), y)}{\partial \theta_2}$ \\
 $\tilde{y}^* = argmin_{\tilde{y}} \, MSE(h, \frac{\partial L(\tilde{f_2}(f_1(x)), \tilde{y})}{\partial \tilde{\theta_2}})$
 \caption{UnSplit: Label Inference}
 \label{alg:label_attack}
\end{algorithm}

\subsection{Label Inference}

\begin{table*}[t!]
    \caption{Mean squared error (MSE) values for the original and estimated inputs, averaged over 5 sets of 10 inputs each, obtained when the attack is performed against a randomly initialized client model (before train) and a client model trained for 20 epochs (after train), and the clone model's classification accuracy on the test sets when the attack is performed after the training phase. Clone model accuracy for the before-train scenario is meaningless as there is no client model to steal (it behaves randomly). The reference accuracy values correspond to the original client model's classification accuracy on the test set.\\}
    \centering 
    \begin{tabular}{c|ccc|ccc|ccc}
        \toprule
        & \multicolumn{3}{c|}{MNIST} & \multicolumn{3}{c|}{F-MNIST} & \multicolumn{3}{c}{CIFAR10} \\ \midrule
              & MSE      & MSE      & Clone       & MSE      & MSE      & Clone       & MSE      & MSE      & Clone     \\
        Split & Before   & After    & Acc. \%     & Before   & After    & Acc. \%     & Before   & After    & Acc. \%   \\ 
        Depth & Train    & Train    & (ref: 98)   & Train    & Train    & (ref: 88)   & Train    & Train    & (ref: 71) \\
        \midrule
        1     & 0.070    & 0.048    & 97.45       & 0.154    & 0.084    & 86.11       & 0.056    & 0.051    & 58.03     \\
        2     & 0.093    & 0.076    & 95.69       & 0.186    & 0.197    & 84.34       & 0.057    & 0.065    & 54.02     \\
        3     & 0.099    & 0.065    & 93.75       & 0.196    & 0.177    & 81.24       & 0.128    & 0.084    & 55.15     \\
        4     & 0.105    & 0.124    & 76.27       & 0.189    & 0.119    & 66.17       & 0.093    & 0.096    & 43.69     \\
        5     & 0.108    & 0.095    & 65.27       & 0.207    & 0.167    & 11.54       & 0.098    & 0.111    & 46.75     \\
        6     & 0.106    & 0.098    & 63.3        & 0.207    & 0.152    & 16.12       & 0.102    & 0.089    & 18.54     \\
        \bottomrule
    \end{tabular}
    \label{tab:mse_acc}
\end{table*}

Before launching the label inference attack (Algorithm \ref{alg:label_attack}), the attacker receives the gradient values from the client layer resulting from a single training example during backpropagation. The attacker also knows the input given to the client model as part of the protocol. Figures \ref{fig:splitnn_server_data} and \ref{fig:splitnn_client_label} are potential SplitNN setups in which the server can perform label inference.

To launch the attack, the attacker randomly initializes a model $\tilde{f_2}$ that has the same architecture with the client model $f_2$. The attacker then computes the gradient values  resulting from backpropagation on $\tilde{f_2}$ for each possible label. The label value that produces the closest gradient values to the gradient values received from the client is output as the predicted label. The attacker can then train its clone model with the predicted labels.

To summarize, as displayed by Equation \ref{eq:label_inference} below and Algorithm \ref{alg:label_attack}, the attacker finds the label $\tilde{y}^*$ that minimizes the distance between the gradients computed from the clone model and those received from the client.
\begin{equation} \label{eq:label_inference}
\tilde{y}^* = argmin_{\tilde{y}} \ MSE(\frac{\partial L(f_2(f_1(x)), y)}{\partial \theta_2}, \frac{\partial L(\tilde{f_2}(f_1(x)), \tilde{y})}{\partial \tilde{\theta}_2})
\end{equation}

\subsection{Number of Clients Does Not Matter}

Although for simplicity we explain our attacks over a setup with a single client and a server, they all generalize to any $n$-client setup without any problem. Two observations help explaining this: a) a SplitNN server trains with a single client at any given time; the attack can be launched against each client as they take their turns training, and b) clients continuously update a single set of parameters; an $n$-client setup is in this way identical to a single-client setup with all the data aggregated at that client (i.e., although there are $n$ physically separated client models, they all follow the same updates as if they were one). 

\begin{figure}[h!]
    \centering
    \includegraphics[width=\linewidth]{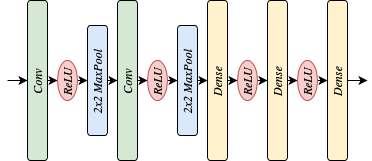}
    \caption{The DNN architecture we used in our experiments for the MNIST and Fashion-MNIST datasets.}
    \label{fig:mnist_model}
\end{figure}
\begin{figure}[h!]
    \centering
    \includegraphics[width=\linewidth]{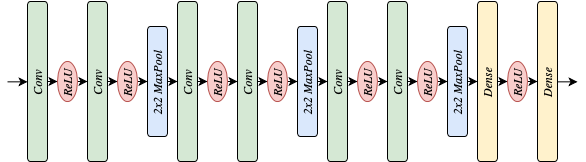}
    \caption{The DNN architecture we used in our experiments for the CIFAR10 dataset.}
    \label{fig:cifar_model}
\end{figure}

\section{Experimental Results}\label{experimental_results}

\subsection{Experimental Setup}

\begin{table*}[h!]
    \caption{Estimated inputs before and after the training phase for different split layers and the MNIST, F-MNIST, and CIFAR10 datasets. The first rows (Ref.) display the actual inputs, and the following rows display the estimates for different split depths as denoted in the Depth column.\\}
    \label{tab:main_results}
    
    \centering
    \begin{tabular}{ccc}
        \toprule
        Depth & Before Training & After Training \\ \midrule
        Ref. & \includegraphics[width=0.42\textwidth]{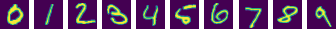} & \includegraphics[width=0.42\textwidth]{results/mnist/targets.png} \\ 
        1    & \includegraphics[width=0.42\textwidth]{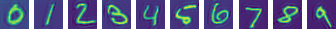} & \includegraphics[width=0.42\textwidth]{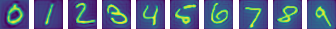} \\
        2    & \includegraphics[width=0.42\textwidth]{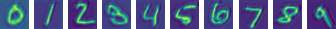} & \includegraphics[width=0.42\textwidth]{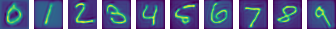} \\ 
        3    & \includegraphics[width=0.42\textwidth]{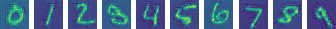} & \includegraphics[width=0.42\textwidth]{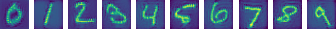} \\ 
        4    & \includegraphics[width=0.42\textwidth]{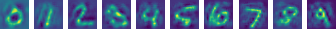} & \includegraphics[width=0.42\textwidth]{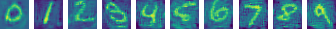} \\ 
        5    & \includegraphics[width=0.42\textwidth]{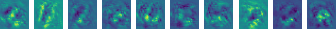} & \includegraphics[width=0.42\textwidth]{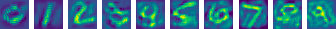} \\ 
        6    & \includegraphics[width=0.42\textwidth]{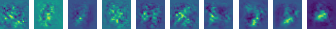} & \includegraphics[width=0.42\textwidth]{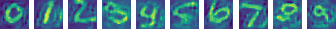} \\ 
        \midrule
        Ref. & \includegraphics[width=0.42\textwidth]{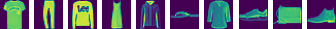} & \includegraphics[width=0.42\textwidth]{results/f_mnist/targets.png} \\ 
        1    & \includegraphics[width=0.42\textwidth]{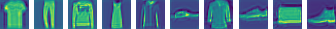} & \includegraphics[width=0.42\textwidth]{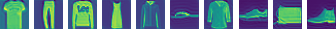} \\
        2    & \includegraphics[width=0.42\textwidth]{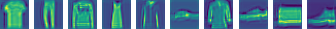} & \includegraphics[width=0.42\textwidth]{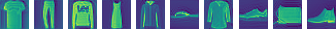} \\ 
        3    & \includegraphics[width=0.42\textwidth]{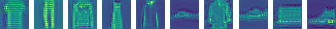} & \includegraphics[width=0.42\textwidth]{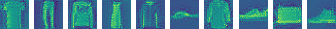} \\ 
        4    & \includegraphics[width=0.42\textwidth]{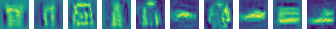} & \includegraphics[width=0.42\textwidth]{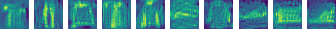} \\ 
        5    & \includegraphics[width=0.42\textwidth]{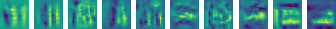} & \includegraphics[width=0.42\textwidth]{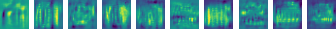} \\ 
        6    & \includegraphics[width=0.42\textwidth]{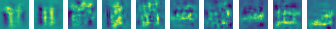} & \includegraphics[width=0.42\textwidth]{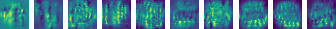} \\ 
        \midrule
        Ref. & \includegraphics[width=0.42\textwidth]{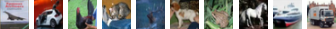} & \includegraphics[width=0.42\textwidth]{results/cifar/targets.png} \\ 
        1    & \includegraphics[width=0.42\textwidth]{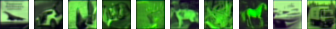} & \includegraphics[width=0.42\textwidth]{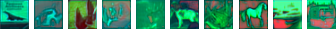} \\ 
        2    & \includegraphics[width=0.42\textwidth]{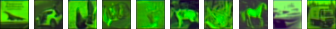} & \includegraphics[width=0.42\textwidth]{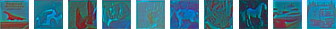} \\ 
        3    & \includegraphics[width=0.42\textwidth]{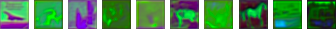} & \includegraphics[width=0.42\textwidth]{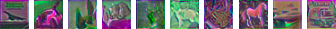} \\ 
        4    & \includegraphics[width=0.42\textwidth]{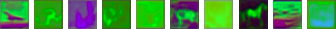} & \includegraphics[width=0.42\textwidth]{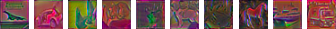} \\ 
        5    & \includegraphics[width=0.42\textwidth]{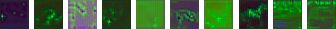} & \includegraphics[width=0.42\textwidth]{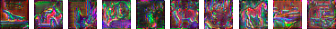} \\ 
        6    & \includegraphics[width=0.42\textwidth]{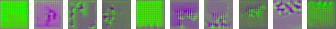} & \includegraphics[width=0.42\textwidth]{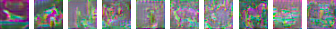} \\ 
        \bottomrule
    \end{tabular}

\end{table*}

\begin{figure*}[t!]
    \centering
    \includegraphics[width=0.95\textwidth]{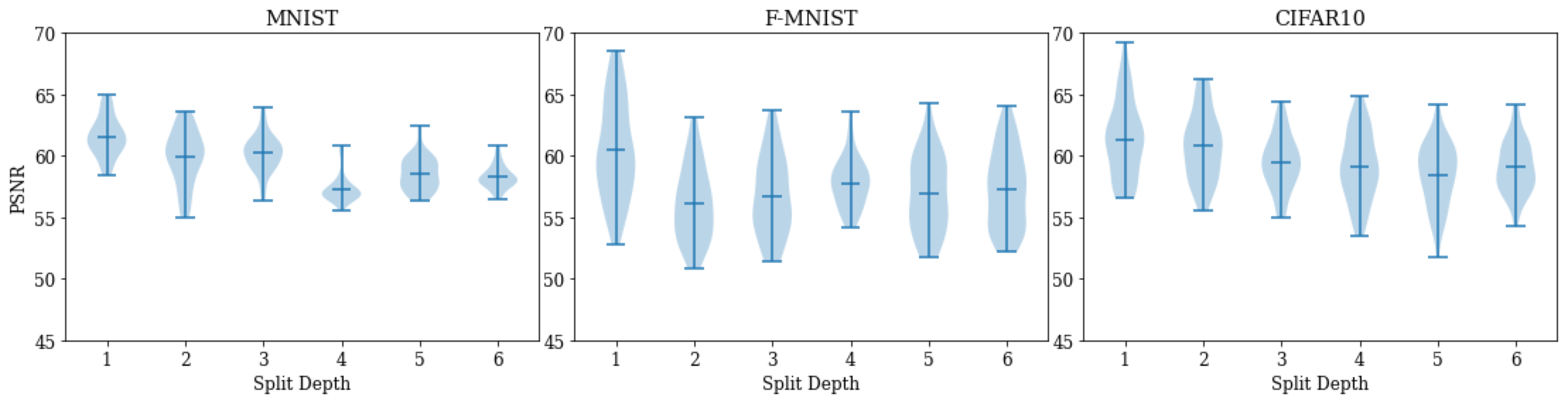}
    \caption{Distribution of PSNR values of the recovered inputs against trained client models, averaged over 5 different target sets of 10 images each for each dataset. The horizontal axis in each plot represents the client model's depth, and the horizontal bars within the plots correspond to mean values.}
    \label{fig:us_psnr}
\end{figure*}

\begin{figure*}[t!]
    \centering
    \includegraphics[width=\textwidth]{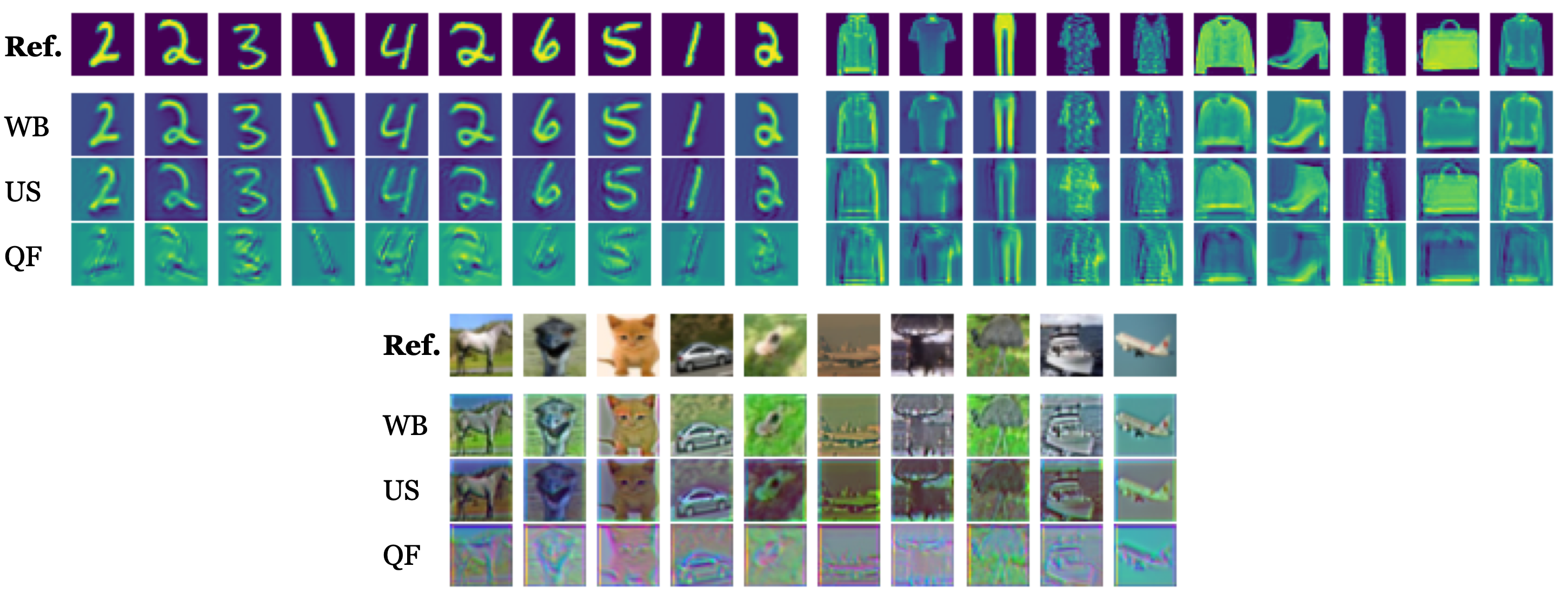}
    \caption{Results obtained over 10 randomly selected images from each dataset (clockwise MNIST, F-MNIST, and CIFAR10) against a trained client model with a split depth of two. The WB and QF rows correspond to the white-box and query-free attacks in \cite{he_model_2019}; the US row corresponds to UnSplit.}
    \label{fig:visual_comp}
\end{figure*}

We first test our attacks, comparing them with related attacks and also against the defensive mechanism described previously, with widely-used image classification benchmark datasets: MNIST \citep{lecun2010mnist}, Fashion-MNIST \citep{xiao2017/online}, and CIFAR10  \citep{Krizhevsky09learningmultiple}. Testing against the distance correlation defense, we limit ourselves to MNIST and Fashion-MNIST due to the significant training overhead caused by the DCOR term in larger models. We use various models consisting of several convolutional and dense layers, as shown in Figures \ref{fig:mnist_model} and \ref{fig:cifar_model}.

We train the original client model using the entire training partition of the datasets, and test the clone model's performance using their test partitions. We perform no post-processing on the estimated inputs. For the sake of brevity, and taking into account that late splits defy the efficient outsourcing purpose of SplitNN, we conduct the experiments for the first six possible layer splits.

For the model inversion loss function (Equation \ref{eq:input_objective}), we set the TV coefficient $\lambda$ to be 0.1 for the first three split layers, and 1 for the rest. We use the Adam optimizer \citep{kingma_adam_2017} with a learning rate of $0.001$ to perform the gradient descent updates. 

Finally, we obtain our results over 5 randomly chosen, distinct image sets of 10 images each, and average the results of those 5 sets.

We implement the attack in Python (v3.7) using the PyTorch library (v1.7.1) \citep{NEURIPS2019_9015}. The time to invert a single input ranged between one and five minutes using a personal computer (2.9 GHz Intel i7 CPUs).

\subsection{Results}

\textbf{Model Inversion \& Stealing.} Figure \ref{fig:visual_comp} (row US) displays the estimated inputs obtained from the model inversion and stealing attack with a split depth of two. Table \ref{tab:mse_acc} displays the MSE values between the original and estimated inputs, as well as the classification accuracy of the clone model, and Figure \ref{fig:us_psnr} displays the PSNR values for each dataset corresponding to different split depths. When the client model is trained, the attacker estimates inputs with reconstruction errors of 0.084, 0.149, 0.083 on average for MNIST, Fashion-MNIST, and CIFAR10 datasets. Against an untrained client model, the error values increase to 0.097, 0.189, 0.089, implying that a trained model is more vulnerable to an attack compared to an untrained model. Furthermore, especially for early splits, the clone model performs very close to the original model on previously unseen data on MNIST, F-MNIST, and CIFAR10. Averaging over the first three splits, the clone model achieves a test classification accuracy of 95.63\% for MNIST, 83.90\% for Fashion-MNIST, and 55.73\% for CIFAR10.


    

\textbf{Effect of training state.} We can infer from Table \ref{tab:mse_acc} that the quality of the estimated inputs is higher when the attack is performed against a trained client model. This is not surprising since a trained model's output preserves more information about the inputs compared to a random, untrained model. However, more importantly, it is misleading to think that an untrained (i.e. randomly initialized) model is not vulnerable to the attack. 
As the results displayed in Table \ref{tab:main_results} (Appendix) demonstrate, an untrained model can leak considerable information as well.

\textbf{Label Inference.} We observe that under the assumption of a client computing only the last layer, aiming to hide the labels from the server while delegating as much work as possible, the attacker can infer the labels with perfect accuracy. After successfully inferring the labels, the attacker can then train its clone model and obtain a model that performs as well as the client model, since they basically follow the same training procedure. Thus, if the client part of the network "protecting" the labels is one layer deep, it does not achieve its purpose.

\begin{figure*}[t!]
    \centering
    \includegraphics[width=0.95\textwidth]{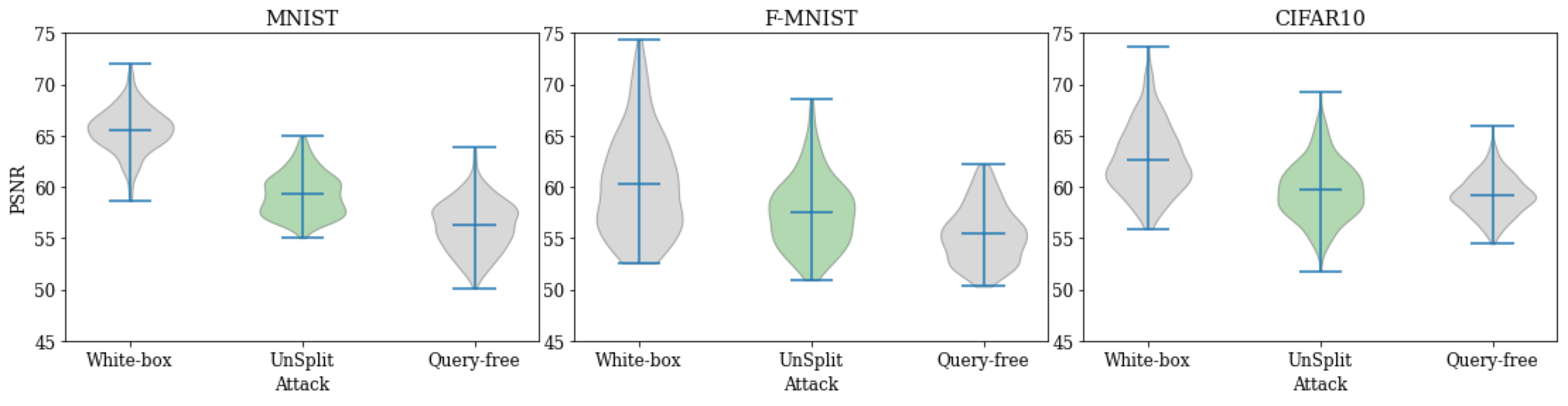}
    \caption{PSNR values of the recovered images obtained against 5 randomly selected target sets of 10 images each for each dataset, averaged over six split layers against a trained model, comparing our work with the white-box and query-free attacks in \cite{he_model_2019}. The horizontal lines in the middle of each plot represent mean values.}
    \label{fig:comp_psnr}
\end{figure*}

\begin{table*}[t!]
    \centering
    \begin{tabular}{l|ccc|ccc}
    \toprule
                    &  \multicolumn{3}{c|}{MNIST}                 & \multicolumn{3}{c}{F-MNIST} \\ \midrule
        $\alpha_1$  & MSE   & Client Acc. (\%) & Clone Acc. (\%) &  MSE    & Client Acc. (\%) & Clone Acc. (\%) \\ \midrule
        0.1         & 0.083 & 97.42            & 98.50           &  0.181  & 84.14            & 84.05 \\
        1           & 0.078 & 96.34            & 96.35           &  0.181  & 82.28            & 82.65 \\
     \bottomrule
    \end{tabular}
    \caption{MSE values of the recovered inputs, and the clone models' test classification accuracy against the DCOR defense. $\alpha_1$ is again the security parameter, denoting the impact of the DCOR term on the loss function.}
    \label{tab:nopeek_mse}
\end{table*}

\begin{figure}[t!]
    \centering
    \includegraphics[width=\linewidth]{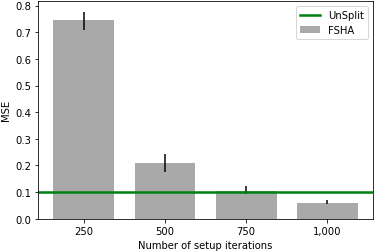}
    
    \caption{Comparison of UnSplit with FSHA as described in \citep{pasquini_unleashing_2021}. The figure displays the MSE values between the estimated inputs and the original values averaged over possible split layers. The vertical lines represent the upper and lower quartiles of the values, while the bars correspond to the means.}
    \label{fig:fsha_compare}
\end{figure}

\begin{table}[t!]
    \caption{Number of black box queries needed for the black box attack scenario in \citep{he_model_2019} to obtain the same PSNR values as UnSplit, averaged over 5 different target sets.\\}

    \centering
    \tabcolsep=0.1cm
    \begin{tabular}{cccc}
    \toprule
    Split & \multicolumn{3}{c}{Number of Black Box Queries} \\ \cmidrule(r){2-4}
    Depth & MNIST & F-MNIST & CIFAR10\\ \midrule
    1     & 12  & 17        & 42     \\ 
    4     & 70  & 76        & 71     \\ 
    \bottomrule
    \end{tabular}
    
    \label{tab:black_box_query}
\end{table}

\textbf{Comparing with other attacks.} Figure \ref{fig:visual_comp} presents the randomly chosen visual results for a split depth of two resulting from UnSplit and the white-box and query-free attacks described in \cite{he_model_2019}; Figure \ref{fig:comp_psnr} displays the PSNR between the recovered images and the original images in the three attacks, averaged over 5 runs each. To compare the attacks under similar threat models, we assume that the attacker does not have access to any specific dataset in any of the scenarios. The estimates generated by the white-box attack are expectedly more similar to the original inputs, and produce higher PSNR values since a white-box adversary has unlimited access to the client model, an unrealistic scenario for a SplitNN setup. On the other hand, UnSplit results in more accurate estimates compared to the query-free attack. 

Table \ref{tab:black_box_query} displays the average number of black-box queries made by the server in \cite{he_model_2019}'s black-box attack to reach the same PSNR values UnSplit obtains. With each query corresponding to a single input (image) given to the client model, the results indicate that if the attacker has a very limited query budget (e.g. 12 for MNIST with a split depth of 1, and 71 for CIFAR10 with depth 4), the black-box attack produces similar results with UnSplit. However, taking into account the details of the default SplitNN setup (Figure \ref{fig:splitnn_client_data}), it is not possible for the server to send queries to the client without violating the protocol. Therefore, black-box attack is not an honest-but-curious attacker model and does not fit our threat model.

\begin{figure*}[t!]
    \centering
    \includegraphics[width=0.8\textwidth]{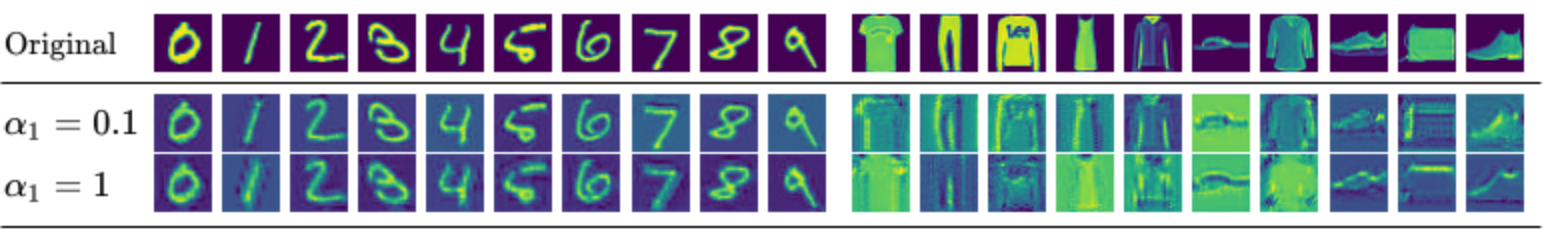}
    \caption{Obtained inputs against the distance correlation defense using the MNIST and Fashion-MNIST datasets. The hyperparameter $\alpha_1$ controls the impact of the DCOR term on the overall loss function, with higher values corresponding to more impact.}
    \label{fig:nopeek_images}
\end{figure*}

Figure \ref{fig:fsha_compare} displays the results of a comparison between UnSplit and FSHA \citep{pasquini_unleashing_2021} on the MNIST dataset. UnSplit performs comparably to FSHA until the FSHA adversary performs around 1,000 setup iterations. Note that the FSHA adversary is stronger, with access to a dataset similar to the training set. FSHA becomes infeasible without such a dataset.

\begin{figure}[h!]
    \centering
    \includegraphics[width=\linewidth]{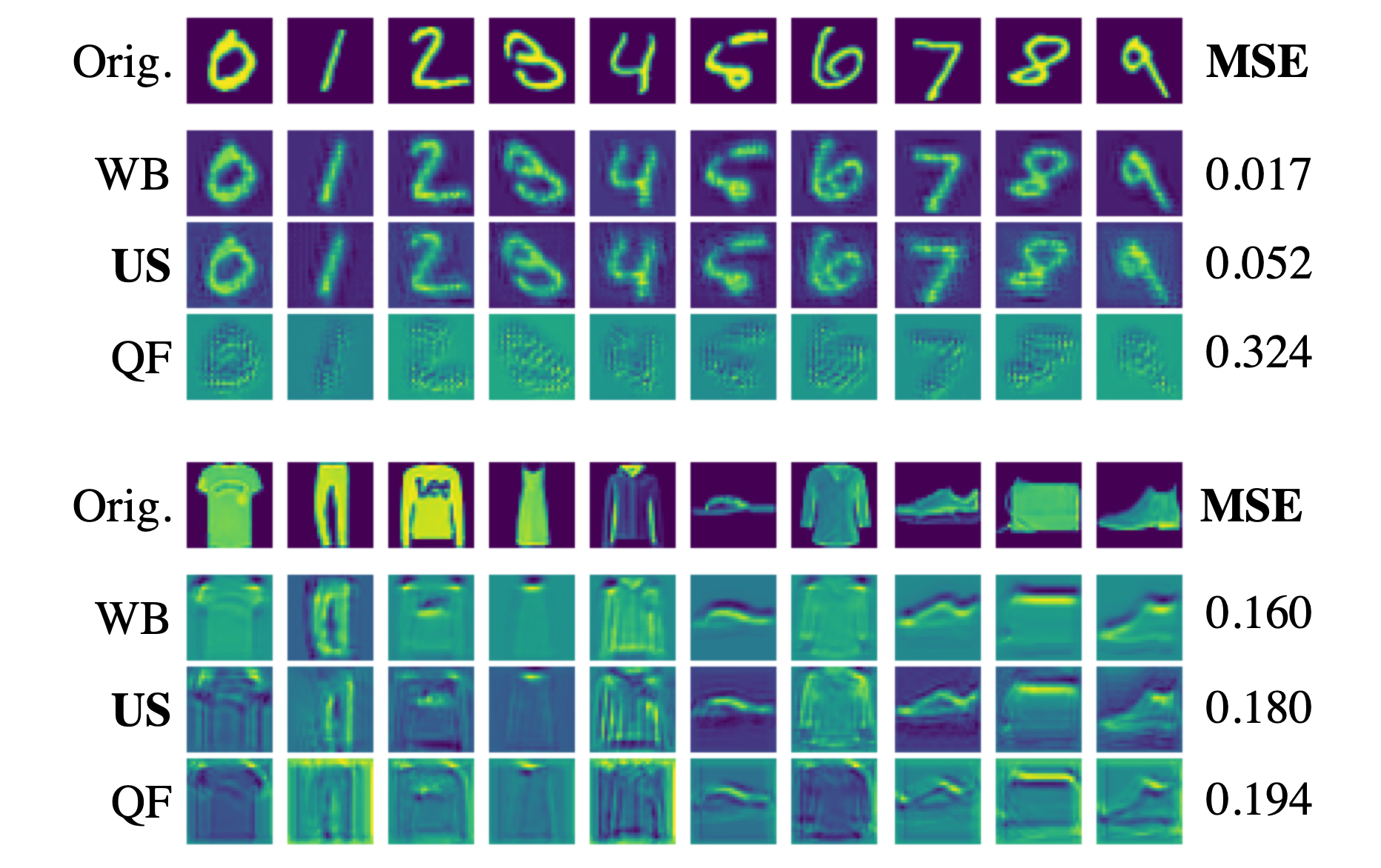}
    \caption{Visual results and the corresponding MSE values against the DCOR defense with $\alpha_1=1$ comparing our attack with those in \cite{he_model_2019} with a split depth of 3, using the MNIST and F-MNIST datasets.}
    \label{fig:dcor}
\end{figure}




\subsection{Results Against the DCOR Defense}
Figure \ref{fig:nopeek_images} and Table \ref{tab:nopeek_mse} display the results of the model inversion \& stealing attack against a client trained with the DCOR defense for 10 epochs, with a client-side model 3 layers deep. Unsurprisingly, the MSE values between the recovered and original images ($\approx$ 0.8 and 0.18 for MNIST and F-MNIST) are higher than the same values obtained without a defensive mechanism (Table \ref{tab:mse_acc} with split depth 3). However, the recovered images are still of high quality compared with the originals, which indicates that the information leakage has not been adequately minimized. Table \ref{tab:nopeek_mse} also indicates that the defensive mechanism provides no protection against model stealing, with the clone model performing as well as (with slight perturbations due to randomness) the target client model. 

\revised{Figure \ref{fig:dcor} displays the visual results of UnSplit and the two other attacks in \cite{he_model_2019} along with their MSE values for the MNIST and F-MNIST datasets. Visually comparing with the results in Figure \ref{fig:visual_comp}, only the query-free attack in \cite{he_model_2019} is heavily impacted by the DCOR defense. This could be because the intermediate outputs supposedly contain a high level information for the model to learn the task, even when a defense mechanism is used. The bottom line is that while DCOR reduces the quality of the inferred inputs, its effect can be substantially reduced with additional capabilities for the attacker (e.g., knowledge of the client model architecture as in our work, or straight white-box access as in \cite{he_model_2019}). }

\section{Conclusion}

Our attacks demonstrate that with the knowledge of the client's DNN architecture alone, it is possible for a honest-but-curious SplitNN server to obtain the inputs given to the model, and a model that performs similarly to the original client model. Furthermore, under the assumption that the final client split has a depth of one, the server can infer the labels with perfect accuracy. These attacks considered together effectively "unsplit" the split learning approach.
Thus, it is of critical importance to warn against such allegedly secure yet blatantly insecure uses of the SplitNN protocol.
\revised{As a testament to this, two recent preprints which follow our work with stronger threat models have been released \cite{liu2022clustering, kariyappa2021gradient}. For example, \cite{kariyappa2021gradient}'s adversary owns the inputs and infers the label-owner's labels (similar to the setup in Figure \ref{fig:splitnn_server_data}). In \cite{liu2022clustering}, the adversary has the list of labels, and one labeled sample for each label. These are both stronger than our threat model as our adversary \textit{only} has access to the backpropagated gradients. }

For the model inversion and stealing attack, its effectiveness decreases as the split layer becomes deeper. This is not surprising since the earlier layers of a DNN contain more information about the inputs. This introduces a performance/security trade-off for the clients. If the data being fed into the DNN is sensitive (e.g. patient data in a clinic), then the data holders can increase the security of the protocol by essentially spending more computing power. 

However, even though expanding more computing resources by way of computing more layers \textit{increases} the security of the protocol, it does not \textit{guarantee} it. Additional mechanisms such as homomorphic encryption are required to provide provable security guarantees. The possibility of our attack under a limited threat model exposes the inherent insecurity of vanilla SplitNN, and highlights the importance of such additional measures to yield the protocol secure.


\section*{Acknowledgements}
We acknowledge support from T\"{U}B\.{I}TAK, the Scientific and Technological Research Council of Turkey, under project number 119E088. 

\balance
\bibliography{references}
\bibliographystyle{ACM-Reference-Format}

\end{document}